\documentclass[aps]{revtex4}

\usepackage{amsmath}
\usepackage{amssymb}
\usepackage{amsfonts}
\usepackage{float}
\usepackage{graphicx}
\usepackage{pstricks}
\usepackage{xcolor}

\begin{document}
\title{Trees and Forests in Nuclear Physics}
\author{M. Carnini}
\email{marco.carnini@features-analytics.com}
\affiliation{Features Analytics, Rue de Charleroi 2, 1400 Nivelles, Belgium}
\author{A. Pastore}
\email{alessandro.pastore@york.ac.uk}
\affiliation{Department of Physics, University of York, Heslington, York, Y010 5DD, UK}

\begin{abstract}
We present a simple introduction to the decision tree algorithm using some examples from nuclear physics. We show how to improve the accuracy of the classical liquid drop nuclear mass model by performing Feature Engineering with a decision tree. Finally, we apply the method to the Duflo-Zuker model showing that, despite their simplicity, decision trees are capable of improving the description of nuclear masses using a limited number of free parameters.
\end{abstract}


\date{\today}

\maketitle

\section{Introduction}

In recent years, there has been an enormous growth  of new statistical tools for data science~\cite{elsl,mehta2019high}. Although these methods are extremely powerful to understand complex data and detect novel patterns, they are still rarely adopted by the nuclear physics community. Only a few groups are currently pioneering the applications of these methods to the field. These topics have been recently discussed in a series of workshops on Information and Statistics in Nuclear Experiment and Theory (ISNET). Recent developments in this field are documented in the associated focus issue published in Journal of Physics G~\cite{Ireland_2015}.  The aim of this guide is to illustrate an algorithm used widely in data analysis. Similarly to our previous guide on bootstrap techniques~\cite{pas19}, we present the decision tree starting from very basic models, then finally apply it to more realistic problems, like improving models for nuclear mass predictions. 

Decision trees are already implemented within major experimental collaborations, such as MiniBooNE, to improve the performances of particle detectors~\cite{tree1,tree2}, but they are not yet widely used in low energy nuclear physics, where they could help to analyse both experimental data~\cite{bailey2017new} and theoretical models.

Following the notation and terminology of Leo Breiman's paper \emph{Statistical Modeling: The Two Cultures}~\cite{breiman}, we want to investigate a process $f$ that transforms some input $X$ into an output $Y$. That is to say, $f$ is a function:
\begin{equation}
f: X \rightarrow Y\;,
\end{equation}
\noindent where the input $X$ can be quite general, from images to a table of data, while the output $Y$ can be a discrete or continuous set. In the first case we speak of a \emph{classification} problem, in the latter of a \emph{regression} problem. 

Rather than focusing on investigating the fine details of the process $f$ with many restrictive assumptions (an approach that is named \emph{data model culture} in \cite{breiman}), we consider $f$ as a \emph{black box} mapping $X$ to $Y$, and we try to approximate it. That is, we give up trying to investigate all the fine details of $f$ and we focus on finding a \emph{representation} (or approximation) $\tilde{f}$ for $f$. $\tilde{f}$ is called a model and it is a function with the same domain $X$ of the process $f$ and codomain $\hat{Y}$:
\begin{equation}
\tilde{f}: X \rightarrow \hat{Y}\;.
\end{equation}

\noindent The process $\tilde{f}$ depends on variables (usually named $features$), parameters (coefficients that can be learned with the algorithm) and $hyper$-$parameters$, that are set before training the model (and thus are not learned).  We will present an extended discussion on how to select the features of the model to improve performances in Sec.~\ref{CART:python}. Another goal for the feature selection process is to reach a representation as parsimonious as possible. 

Since ``all models are wrong, but some are useful"~\cite{box1976science}, it is necessary to introduce a definition of what a good model looks like in order to pick the best one out of a set of possible candidates. Or in other terms, we need to assess how faithfully $\tilde{f}$ represents the process $f$. Mathematically, the goal for training a model $\tilde{f}$ is to minimise a particular \emph{scoring function}, sometimes improperly called ``a metric''. Without loss of generality, we are considering only the minimisation problem: changing the sign of a scoring function to be maximised reduces the problem to a minimisation task. 

For example, a natural choice for the scoring function is the mean squared error (MSE) or variance, that is the difference between the predicted value ($\hat{Y}$) and the observed, experimental data ($Y$)~\cite{pedregosa2011scikit}:
\begin{equation}\label{MSE}
\text{MSE}(Y, \hat{Y}) = \frac{1}{N} \sum_{i=1}^{N}(Y_i-\hat{Y}_i)^2\;.
\end{equation}
\noindent It is worth noting that this is not the only option, and within the Machine Learning literature we encounter other scoring functions as the logarithmic mean squared error (MSLE):

\begin{equation}
\text{MSLE}(Y, \hat{Y}) = \frac{1}{N} \sum_{i=1}^{N} (\log_e (1 + {Y}_i) - \log_e (1 + \hat{Y}_i) )^2\;,
\end{equation}

\noindent or  the median absolute error \cite{pedregosa2011scikit}:
\begin{equation}
\text{MedAE}(Y, \hat{Y}) = \text{median}(\mid Y_1 - \hat{Y}_1 \mid, \ldots, \mid Y_N - \hat{Y}_N \mid).
\end{equation}

Different scoring functions correspond to different modelling choices and the importance we assign to specific sub-sets of the database. The use of MedAE would be more appropriate to obtain a model that is robust to outliers: a few poorly described experimental points will not alter significantly the performances. In the current work, we have chosen the mean standard error (or equivalently its square root, RMS) which is the default in most libraries. Given the high accuracy of measurements in nuclear physics, especially for masses as discussed here, we do not need to worry about possible outliers in our data sets and MSE therefore represents a reasonable choice.

Another important aspect in building a model is the decision on the tradeoff required between model performances and \emph{explainability}. That is, the choice between (possibly) better performances with the chosen scoring and easier explanations of the model in plain language. 
Among the regressors usually considered to be explainable are linear regression and decision trees. However, some recent research allows explaining (although approximately) even the results from algorithms deemed black-box, such as neural networks, or such as gradient boosting in explainable models like linear regression~\cite{ribeiro2016} and simple decision trees~\cite{boz2002}. 

In this guide, we chose to illustrate decision trees because they retain explainability, they do not rely on the assumption of linearity nor on the linear independence of the \emph{features}, and they are not significantly affected by monotonic transformations (no input data scaling is required, nor monotonic transformations like taking the logarithm or the square of one variable). Also, decision trees are the key elements in building other regressors like Random Forests~\cite{breiman2001} or Xgboost~\cite{chen2016} that usually perform better with regard to scoring.

Last but not least, an important aspect of modelling is the balance between the complexity of the chosen model and the generality of the results. As an analogy,  it is useful to consider the problem of approximating $N$ experimental distinct points using a polynomial. A complex polynomial of degree $N$ will be able to describe perfectly the data. Whenever some new data are added, the perfect description will (in general) no longer be true. A correct assessment of the performance of a regressor should be performed on unseen data, \emph{i.e}. data that were not used during the training.

A common practice to estimate the performance on unseen data is the $k$-fold cross validation, with $k \in \mathbb{N}$. In essence, the data are permuted and then separated in sets of size $k$ with each subset (fold) roughly of the same cardinality. The model is then trained on $k-1$ subsets and validated on the subset not used while training. As an extreme case, when $k=N-1$, all data but one are used for training and the performances are assessed on only one datum. This scheme is called ``leave-one out validation''~\cite{lachenbruch1968}.

In the following sections, we will illustrate the behaviour of decision trees using some nuclear mass models. The article is organised as follows: in Sec.\ref{sec:tree}, we provide an introduction to what a decision tree is, using very simple examples. In Sec.\ref{nuc:m:model}, we introduce the nuclear models to which we will apply the decision trees. The results of our work are presented in Sec.\ref{sec:conc} and we illustrate our conclusions in Sec.\ref{sec:conc}.
In the Supplementary Material, we provide the Python script used to perform the calculations. The script has been structured in the same way as the current guide to facilitate its usage~\footnote{We provide an HTML version of the material at the web address  https://mlnp-code.github.io/mlnp/}.

\section{Decision Tree}\label{sec:tree}

\noindent With a decision tree, the function $f:X \rightarrow Y$ is approximated with a step function with $n$ steps as

\begin{equation}\label{eq:step}
\tilde{f} = \sum_{i=1}^n \alpha_i \mathbb{I}(\Omega_i)\;,
\end{equation}

\noindent with $\Omega_i \subseteq X$, $X\subset \mathbb{R}^d$  where $d$ is the number of features, $\mathbb{I}(\Omega_i)$ is the indicator function:

\begin{equation}
\mathbb{I}(\Omega_i) = \begin{cases}
1 & x\in \Omega_i \\
0 & x\notin \Omega_i \;
\end{cases}
\end{equation}

\noindent and $\Omega_i$ are half-planes in $\mathbb{R}$. 

 Any measurable function can be approximated in terms of step functions \cite{maderna1985lezioni}, thus the approximation is justified as long as the function $f$ is expected to be measurable. That is, using enough step functions we can approximate any measurable function.

Each step function required to build the model $\tilde{f}$ (the tree) is called a \emph{leaf}, thus the number of leaves of the model corresponds to the number of step functions employed.
\noindent In order to determine the optimal values for the $\alpha_i$ and $\Omega_i$ of Eq.\ref{eq:step}, one should provide a splitting criterion; for example, being an extreme value (maximum or minimum) for a given function $\mathcal{L}$. Here, we decide to minimise the $\mathbb{L}^2$ norm of the difference between $f$ and $\tilde{f}$, that is:

\begin{equation}
\mathcal{L} = \|f - \tilde{f}\|_2\;.
\label{eq:l2norm}
\end{equation}

\noindent This function should be chosen to approximate the scoring function selected; then determining the extremes of $\mathcal{L}$ guarantees that we have optimised the desired scoring function.
Since in this guide we chose as a scoring function the MSE, a natural choice for the splitting criterion is the $\mathbb{L}^2$ norm of Eq.\ref{eq:l2norm}; we will use it through all the following examples.

We are going to focus on the CART algorithm as presented in~\cite{CART, elsl}. Calculating all the possible splits for all the features to get the optimal $\tilde{f}$ as in Eq.\ref{eq:step} is computationally unfeasible for large data sets. For this reason, a greedy approximation is used for training the decision tree: at every iteration of the algorithm, a local optimal split is selected. This is a heuristic approach and there is no guarantee of converging to the global optimum.

At the first step of the algorithm, all the possible splits for all the features are explored, and the best split (that minimises $\mathcal{L}$) is selected. Then, all the data are split between the two leaves (a leaf for each half-plane). Then, for every leaf, the procedure is iterated until a stopping criterion is reached. There are many different stopping criteria: a leaf can not be further split if it contains only one observation or if all the features are constant. However, the training is usually stopped as a modelling choice to avoid poor performance on unseen data ($overfitting$): once a given number of leaves or a maximum depth (that is, a maximum number of splits between the root and any leaf) are reached, the algorithm stops. Alternatively, leaves are not split if they contain fewer than a specified number of observations. In this process, some of the features may have never been used; in this case, they are irrelevant to the model, and their absence from the input will make no difference.

In the next subsections, we provide some examples of how a decision tree operates, by showing artificially simple examples of trees with few features and very few leaves that can be easily understood. The more realistic cases will be illustrated in Sec.\ref{nuc:m:model}.

\subsection{A single variable example}\label{subsec:single}

\noindent As a first example, we illustrate the first iteration of a decision tree, splitting the data (a single feature) into only two leaves. We build an example training set $X_{tr}$ defined as the union of two sets $X_1$ and $X_2$. $X_1$ contains random points uniformly distributed in $[0, 0.5]$ and analogously $X_2$ with points in $(0.5,1]$. Notice that $X_1 \cap X_2 = \varnothing$. In this case, there is only one feature, so $d=1$. In this example as well as in the following one, we will directly code the necessary steps to obtain the desired solution; the more advanced reader can skip these two cases to Sec.\ref{CART:python} where an existing library is used.

\begin{figure*}
\begin{center}
\includegraphics[width=0.4\textwidth,angle=0, trim={0 30 0 0}]{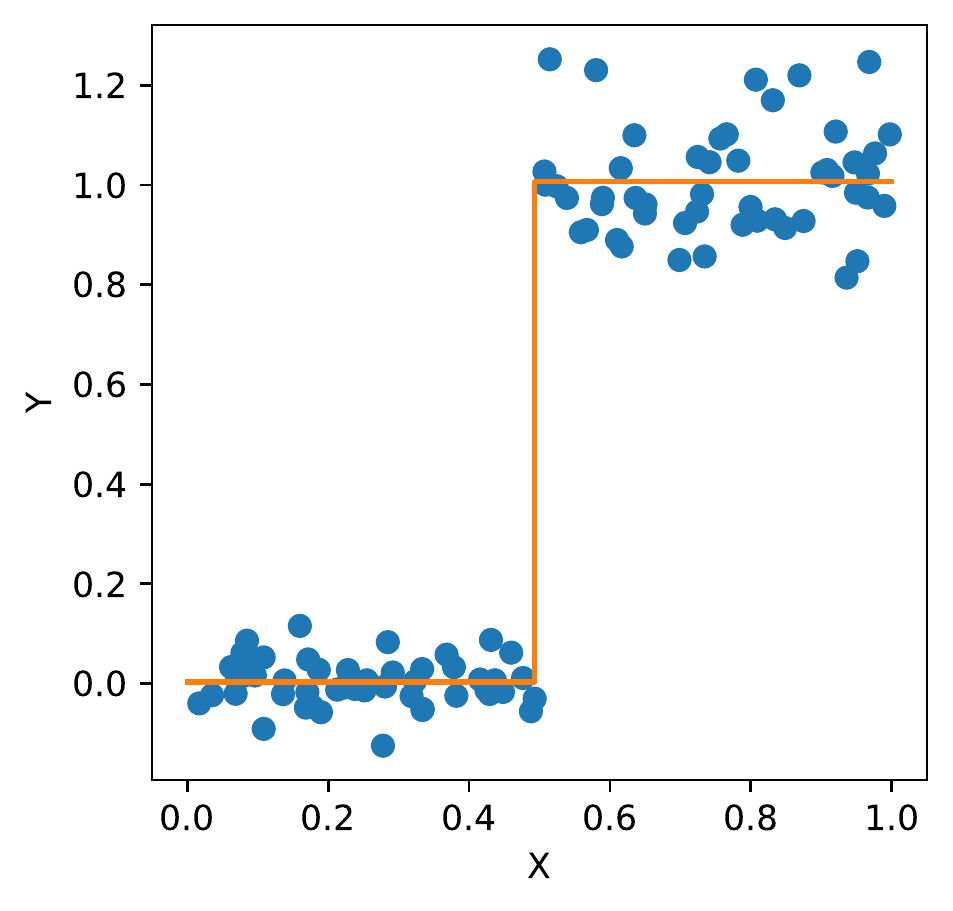}
\end{center}
\caption{The full dots represent the \emph{original} data set, while the line represents the approximating function $\tilde{f}$. See text for details.}
\label{Ex1}
\end{figure*}

For the images through $f$ of $X_1$ ($X_2$), here named $Y_1$ ($Y_2$), we use a Gaussian distribution with mean 0 (1) and a variance of 0.05 (0.1). The training set is illustrated in  Fig.\ref{Ex1}. All figures presented in this article have been realised with the Python~\cite{van1995python} library {\ttfamily matplotlib}~\cite{Hunter:2007}.

 By construction, the data set can be fully described using a decision tree with only two leaves, that is:
\begin{eqnarray}\label{eq:tree:2leaves}
\tilde{f}=\alpha_0 \mathbb{I}(\Omega_1) + \alpha_1 \mathbb{I}(\Omega_2)\,.
\end{eqnarray}

\noindent By visually inspecting the data shown in Fig.\ref{Ex1}, we notice that the current data belong to two groups and a single split along the $x$ axis will be enough to describe them.
\noindent In more advanced examples, the number of leaves will be selected algorithmically. 
\noindent To train a decision tree means to determine the function given in Eq.\ref{eq:tree:2leaves}, in such a way that 

\begin{equation}\label{eq:L2norm}
\mathcal{L} = \|f - \alpha_0 \mathbb{I}(\Omega_1) - \alpha_1 \mathbb{I}(\Omega_2)\|_2\;,
\end{equation}

\noindent is minimal. $\mathcal{L}$ is equivalent to Eq.\ref{MSE} apart from a global scaling factor. The sets $\Omega_1$ and $\Omega_2$ are defined as

\begin{eqnarray*}
\Omega_1 &=& \{x | x\leq x_*\},\\
\Omega_2 &=& \{x | x > x_*\}.
\end{eqnarray*}

\noindent It is easy to prove that the constant which best approximates (in terms of $\mathbb{L}^2$ norm) a set of values is the average of the values, $\bar{x}$.


\begin{figure*}
\begin{center}
\includegraphics[width=0.4\textwidth,angle=0, trim={0 30 0 0}]{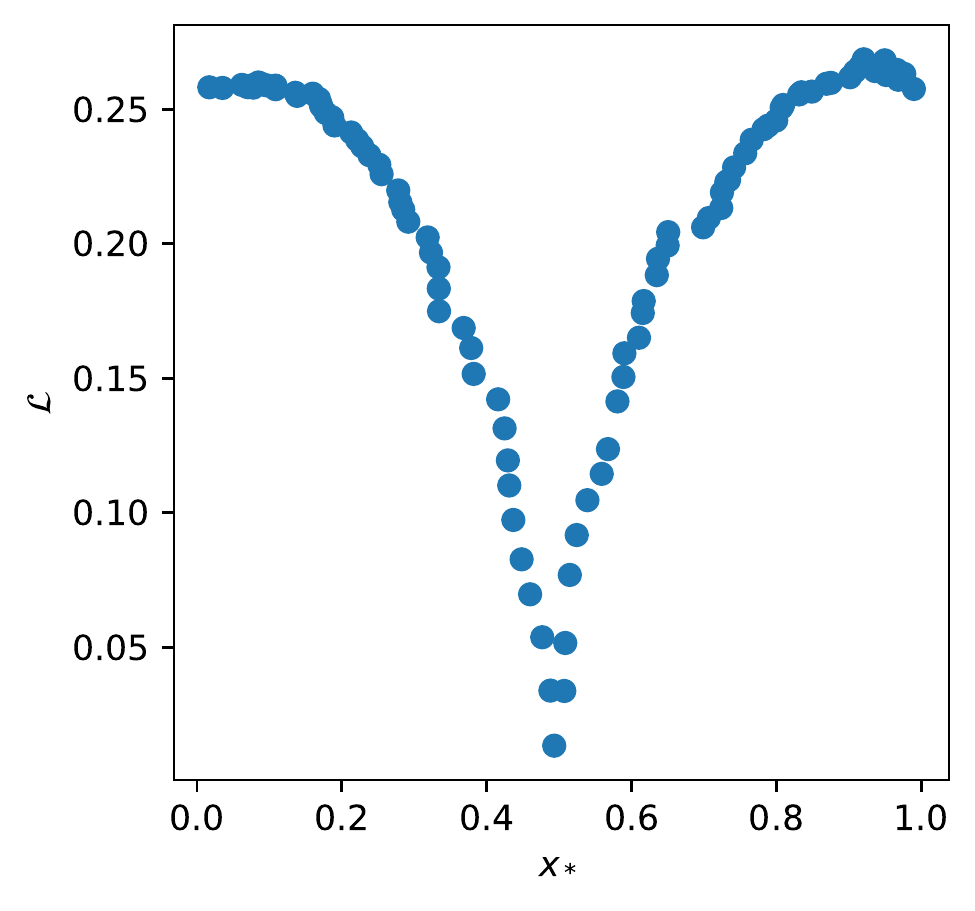}
\end{center}
\caption{Evolution of the $\mathcal{L}$ norm defined in Eq.\ref{eq:L2norm} as a function of the splitting point $x^*$. See text for details.}
\label{var1}
\end{figure*}

\noindent The optimal value for $x_*$ (0.5) is obvious when the generating process for the data is known, but how do we determine it when $f$ is not known? The answer is reported in Fig.\ref{var1}, where we plot $\mathcal{L}$ as a function of $x^*$. The optimal value $x^*$ is the one that  minimises $\mathcal{L}$. %


For this particular case we obtained $x_*=0.493$, $\alpha_0=0.003$ and $\alpha_1=1.007$. More details for reproducing the results are provided in the Supplementary Material.
Thus the decision tree reads:

\begin{equation}
\tilde{f} = \begin{cases}
0.003 & \text{if }\, x \leq 0.493 \\
1.007 & \text{if }\, x > 0.493 \;.\\
\end{cases}
\end{equation}

\noindent Following these simple steps, we have built a model $\tilde{f}$ that is able to provide a reasonable description of the main structure of the data, \emph{i.e.} we recognise that the data are separated in two groups.  This is represented by the solid line in  Fig.\ref{Ex1}. We say that this tree has two \emph{leaves} since we have separated the data into two subgroups. Notice that the MSE is not exactly equal to zero since there was some noise in the generated data.

\subsection{A two variables example}

 We now consider a slightly more complex  problem with two variables $x_1,x_2$. The aim of this example is to familiarise with the concept of multiple splits to treat a complex problem via simple operations. As in the previous case, we will apply the basic steps to explicitly build a decision tree. We generate a new set of data points as:

\begin{equation}\label{eq:ds1}
X = \left\{(x_1, x_2) |x_1 \sim \mathcal{U}(0,1), x_2 \sim \mathcal{U}(0,1)\right\}\;,
\end{equation}

\noindent with response:

\begin{equation}
Y = \begin{cases}
10.0 & \left\{(x_1, x_2) \in X | x_1 \leq 0.5\right\}\;,\\
1.0 & \left\{(x_1, x_2) \in X | x_1 > 0.5, x_2 \leq 0.5\right\}\;, \\
0.0 & \left\{(x_1, x_2) \in X | x_1 > 0.5, x_2 > 0.5\right\} \;.
\end{cases}
\end{equation}

\begin{figure*}
\begin{center}
\includegraphics[width=0.4\textwidth,angle=0, trim={0 30 0 0}]{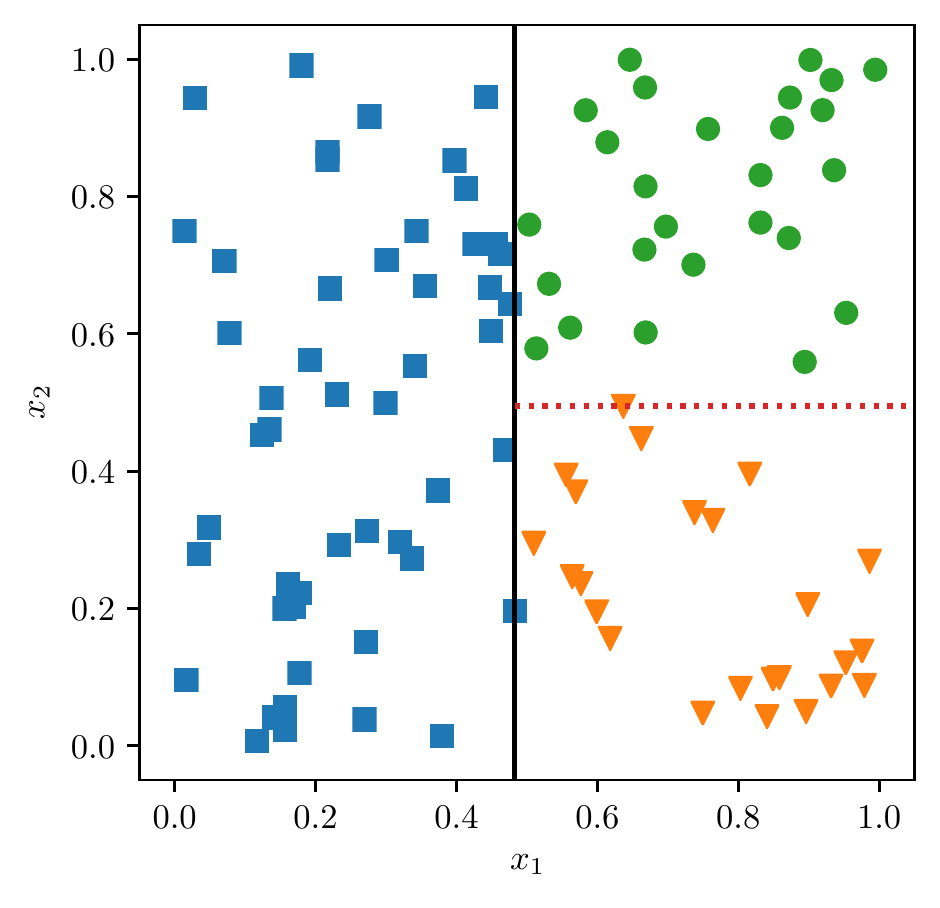}
\end{center}
\caption{Graphical representation of the data set $X$ defined by Eq.\ref{eq:ds1}. The squares correspond to $Y=10$, the dots $Y=1$ and the triangles to $Y=0$. The solid line corresponds to the first optimal split, the dotted line to the second split performed by the decision tree. See text for details.}
\label{fig:tree2v}
\end{figure*}

 Graphically, we represent this data set in Fig.\ref{fig:tree2v}. The data are clearly clustered (by construction) in three regions of the $x_1,x_2$ plane. The aim of the current example is to illustrate how to perform successive splits to correctly identify these regions.
 
 We apply a two-step procedure: firstly, we separate the data along the $x_1$ direction. Following the procedure highlighted in the previous example, we create a model in the $x_1$ direction of the form
 
 \begin{eqnarray*}
 \tilde{f}=\alpha_0 \mathbb{I}(\Omega_1) + \alpha_1 \mathbb{I}(\Omega_2).
 \end{eqnarray*}
 
 \noindent We now perform a systematic calculation of the $\mathcal{L}$ norm looking for the value $x^*$ that leads to its minimal value. We refer to the Supplementary Material for details.
 We find $x_*=0.482$ as the value for dividing the plane. By observing the data, we see that there is no gain by adding further splits in this direction. We will come back to this aspect in the following sections.
We can further refine the model by adding an additional separation along the $x_2$ direction. 
The procedure follows the same steps as before and we find that $x_*=0.495$. The result is reported in Fig.\ref{fig:tree2v}.

An important quantity for analysing the model and for assessing the importance of its input variables is an estimate of the feature importance. This is particularly relevant for the ensemble methods that rely on decision trees: while with a single decision tree the role of the features is obvious once the tree is visually represented (see for example Fig.~\ref{fig:tree2v_loss}), it is unpractical to represent all of the decision trees in a Random Forest (there may be thousands). Also, a feature may participate multiple times in different trees, so a definition of importance should take this into account.

Starting from Fig.~\ref{fig:tree2v_loss}, we want to assess the importance of the features to the building of the decision tree.
We recall here that by \emph{features} we mean the variable of the model. In this particular example, we have considered $x_1,x_2$ as a natural choice, but one could also consider other combinations: $x_1,x_2,x_1+x_2,x_1/x_2,\dots$. We refer the reader to Sec.\ref{sec:feature} for a more detailed discussion.

Following Ref.~\cite{pedregosa2011scikit}, we calculate the feature importance in  the following way: for each split $s$ in the tree, we calculate the weighted impurity decrease as

\begin{equation}\label{eq:importance}
\frac{N_s}{N}\left(I - \frac{N_{s,R}}{N_s}I_R- \frac{N_{s, L}}{N_s} I_L\right).
\end{equation}

\noindent $N$ is the total number of observations, $N_s$ is the number of observations at the current node, $N_{s, L}$ is the number of samples in the left leaf, and $N_{s, R}$ is the number of samples in the right leaf. $I$ represents the impurity (in our case, MSE), with the subscripts having the same meaning as before.

By inspecting Fig.\ref{fig:tree2v_loss}, we observe that for the first split there are in the current node (the root of the tree) as many observations as the total, that is $N=N_s=100$. The initial impurity (MSE) is 22.782, $N_{s,R}=N_{s,L}=50$, right impurity is 0, left impurity 0.25. Thus we obtain

$$
\frac{100}{100} \times \left(22.782-\frac{50}{100}\times 0 -\frac{50}{100}\times 0.25 \right) \simeq 22.657.
$$

\noindent For the second split, we get:

$$
\frac{50}{100} \times \left(0.25-\frac{24}{100}\times 0 -\frac{26}{100}\times 0.0 \right) \simeq 0.125.
$$

\noindent Normalising the total weighted variation to 1, we obtain that the column $x_1$ has importance equal to 99.5\% for the model, while $x_2$ has an importance of 0.5\%. If the variables entered in different splits, the relative importance would be summed.

Estimating feature importance is fundamental for improving the quality of the model: by discarding irrelevant features, \emph{i.e.} features that are not reducing the impurity in the tree, more parsimonious models can be trained. This is especially useful for models involving hundreds (or thousands) of features. For example, anticipating the results of the following section, we see that in  Figure \ref{fig:tree:LD3}, a simpler model could be obtained using only 6 features instead of 9. Whenever features are generated for improving the model, a critical assessment of their relevance should be performed.

After these examples that were easily implemented with a few lines of code, in the next sections (and for realistic problems, in terms of the number of features and the number of possible leaves) we are going to rely on existing libraries.

\subsection{A two variables example (revisited)}\label{CART:python}

In this section, we make use of the function {\ttfamily DecisionTreeRegressor} from the Python package {\ttfamily scikit-learn} to determine the structure of the tree, using all default hyper-parameters apart from the number of leaves. For the sake of simplicity, we still consider $x_1,x_2$ as features of the model and we also impose the number of leaves to be three. In more advanced examples, we will let it be a free parameter.

In this case, contrary to the previous example, we do not need to decide if the split along the $x_1$ direction happens before the one along the $x_2$ or vice-versa: all possible splits on the available data for all features are explored with the algorithm.

\begin{figure*}[!h]
\begin{center}
\includegraphics[width=0.28\textwidth,angle=0, trim={0 30 0 0}]{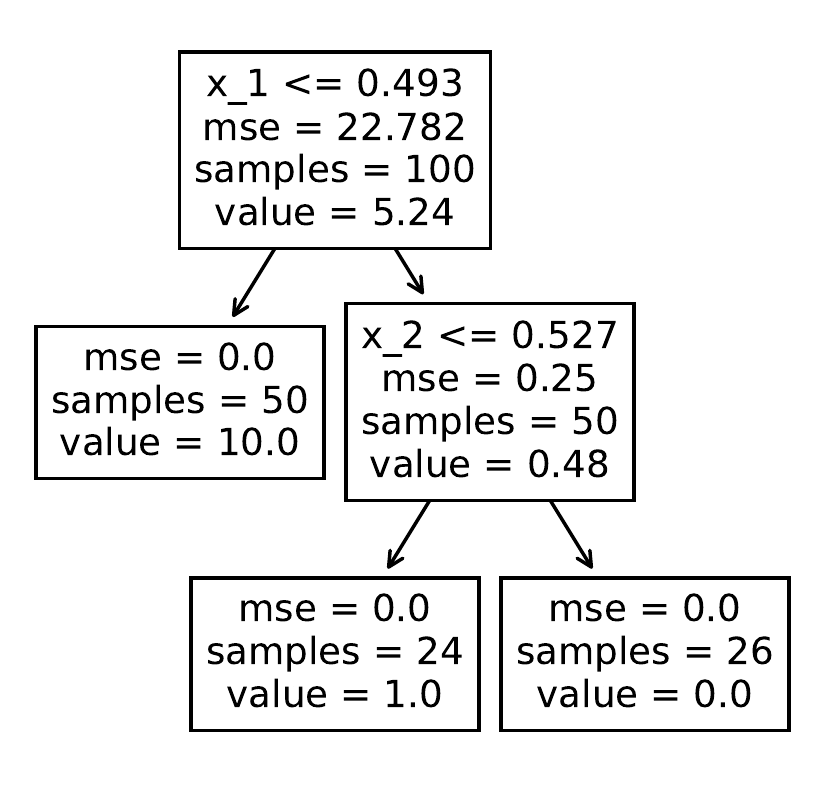}
\end{center}
\caption{Decision tree with two variables obtained using the {\ttfamily scikit-learn} package~\cite{pedregosa2011scikit} for the data set reported in Fig.\ref{fig:tree2v}.}
\label{fig:tree2v_loss}
\end{figure*}

The algorithm used to perform such a split can be  represented as in Fig.\ref{fig:tree2v_loss}  using the {\ttfamily scikit-learn} package~\cite{pedregosa2011scikit}, but an analogous result could have been obtained using R~\cite{rpartplot}, and the libraries {\ttfamily rpart}~\cite{rpart} for model training and {\ttfamily rpart.plot} \cite{rpartplot} for visualisation. 

The visualization of a {\ttfamily scikit-learn} tree consist of a series of boxes counting basic information: the value of the variable at which the separation takes place, but also the mean value of the data (named \emph{value}) and the impurity (MSE in our example). It also provides information concerning the amount of data grouped in each leaf.
In this case, the tree has a total of three leaves. As the MSE is zero (thus, minimal) on each leaf, adding extra splits to the model would not lead to any real gain in the description of the data, but it would only increase the model complexity.


\subsection{The importance of Feature Engineering}\label{sec:feature}

In the previous examples, we have approximated the data using simple step functions; although this choice is mathematically justified, the problem is that the approximation may lead to a single observation per leaf, with the result that the generalisation on unseen data may be unsatisfactory.

To overcome the problem, and possibly to make the models easier to explain, it is important to explore the data and apply \emph{convenient} transformations on the input variable for the model that may highlight some patterns. We consider as an example the case of a two variable data-set obtained as follow:

\begin{equation}
X = \left\{(x_1, x_2) |x_1 \sim \mathcal{U}(0,1), x_2 \sim \mathcal{U}(0,1)\right\}\;,
\end{equation}

\noindent where the response is chosen as

\begin{equation}
Y = \begin{cases}
1.0 & \left\{(x_1, x_2) \in X | x_1^2 +x_2^2 \leq 1\right\}\\
0.0 & \left\{(x_1, x_2) \in X | x_1^2 +x_2^2 > 1\right\}\,. 
\end{cases}
\end{equation}

\begin{figure*}
\begin{center}
\includegraphics[width=0.4\textwidth,angle=0, trim={0 30 0 0}]{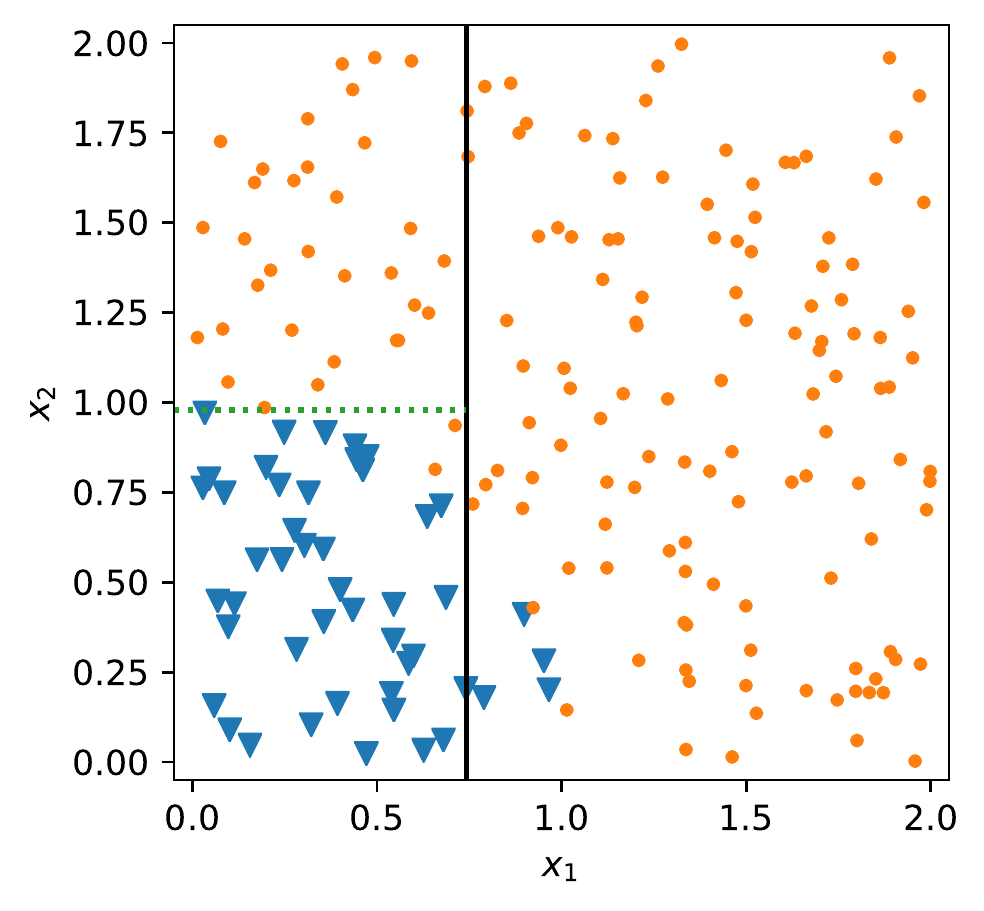}
\includegraphics[width=0.4\textwidth,angle=0, trim={0 30 0 0}]{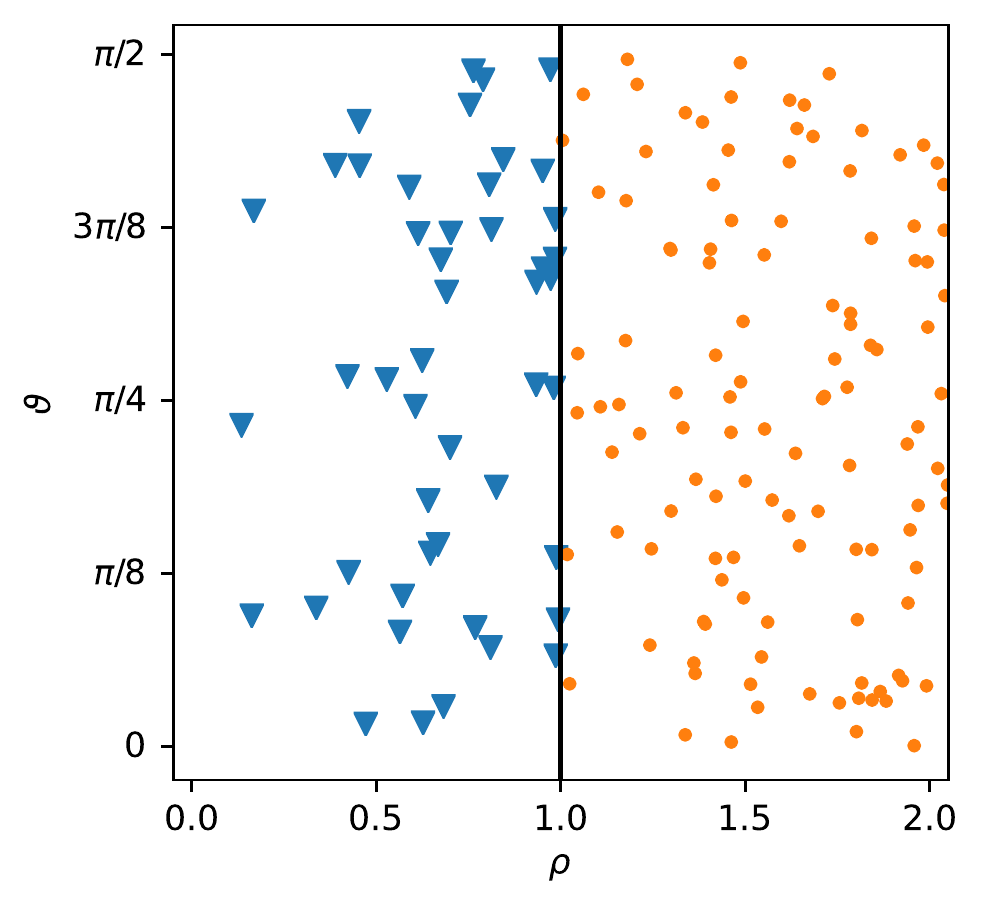}
\end{center}
\caption{Graphical representation of the data set (X,Y): on the left panel the Cartesian coordinates are used while on the right panel the data are represented in polar form. The triangles correspond to $Y=1$ and the dots to $Y=0$. The solid lines represent the cuts done to reproduce the data.}
\label{fig:tree3}
\end{figure*}

\noindent In the left panel of Fig.\ref{fig:tree3}, we illustrate the data points using Cartesian coordinates. 
By using {\ttfamily DecisionTreeRegressor}, we build a series of decision trees as a function of the number of leaves. In Tab.\ref{tab:leaves}, we report the MSE of the tree for various number of leaves.

\begin{table}
\begin{center}
\begin{tabular}{c|c}
\toprule
MSE & leaves\\
\hline
0.278 & 2\\
0.073 & 3\\
0.164 & 4\\
0.042 & 5\\
0.222 & 6\\
0     &7 \\
\hline
\botrule
\end{tabular}
\end{center}
\caption{Evolution of MSE with the number of leaves of the tree for the data set shown in Fig.\ref{fig:tree3}.}
\label{tab:leaves}
\end{table}

From the table, we see that the lowest MSE is given by a 7-leaves tree. This tree is quite complex, with leaves containing only 2 or 4 observations. For illustrative purposes let's consider the case with 3 leaves, which corresponds to $MSE=0.073$.
The splits are performed along the $x_1$ direction at $x_*=0.742$ and  along the $x_2$ direction at $x_*=0.979$. The result is reported using solid and dashed lines in the left panel of Fig.\ref{fig:tree3}.

By inspecting the data, we realise that by applying a unitary transformation from Cartesian to polar coordinates:
\begin{eqnarray}
x_1&=&\rho \sin \theta,\\
x_2&=&\rho\cos \theta,
\end{eqnarray}

\noindent we can highlight a very specific structure in the data. The result of such a transformation is illustrated in the right panel of Fig.\ref{fig:tree3}. Using these new variables in the model obviously helps the performances: using a single split, \emph{i.e.} a tree with two leaves, we obtain a total MSE of zero using fewer leaves and with features that are easier to understand. Only one feature is relevant ($\rho$), so while the possible splits on the other features are explored with the algorithm, they are irrelevant and do not play any role in the model.  Notice that by construction, there is only radial dependence and no dependency on the phase. Thus we obtain a more parsimonious model by applying Features Engineering.

\subsection{Random forests}

We conclude this section on decision trees by introducing the concept of a Random Forest (RF). Following \cite{breiman2001, elsl, louppe2014}, we define a RF as an ensemble of decision trees. The first step for training a RF is to use bagging, also named bootstrap aggregating  \cite{efron1979, breiman1996}. Given a training set $X_{tr}$, a uniform sampling with replacement is performed, obtaining two data sets: one containing the sampled data (with repetitions), $X_1$, and one containing the data that were never sampled, named out-of-the-box (OBB), $X_2$. $X_2$ contains roughly one-third of the initial observations.

In fact, given a number $N$ of observation, assuming no repetitions in the data, the probability for a datum of \emph{not} being extracted at the $i$-th draw is simply $p_i = 1-\frac{1}{N}$. Thus the probability of not being extracted after $N$ draws, having replacement and assuming all the draws to be  independent, is:

\begin{equation}
p = \prod_{i=1}^M p_i = \left(1-\frac{1}{N}\right)^N\underset{N \rightarrow \infty}{\rightarrow}  \frac{1}{\mathrm{e}}\approx \frac{1}{3}\;.
\end{equation}



The second step of the RF algorithms consists of selecting a  random subset of the features of $X_1$. The number of features used is an adjustable hyper-parameter of the algorithm; with \cite{pedregosa2011scikit} for example, the user provides the fraction in $(0, 1)$ out of the total number of features that will be used for each tree. A decision tree is then built on $X_1$ using only the selected features, and the performances are estimated on $X_2$. In \cite{pedregosa2011scikit} the procedure is fully automatic and many trees can be trained in parallel, but the estimation on $X_2$ is not performed by default (but the option can be activated). Repeating the bagging and the tree training for a number $T$ of trees (another adjustable hyper-parameter) and averaging the response $\hat{Y}$ over the ensemble of predictions, a Random Forest is obtained. 

The bagging and the random selection of features are tools to inject noise in the training data. The noise can be reduced by averaging on the response of each tree, and this empirically improves the performance of the regressor~\cite{efron1979, breiman2001}. For the theoretical reasons why the performances are better after injecting some noise in the system, the reference is \cite{breiman2001}. 

Intuitively, the trees in the forest should not all provide the same response, otherwise averaging on all the trees would be of no benefit. Consider for example the dataset $X$ of Eq.~\ref{eq:ds1}: if all the trees are trained on the same data with the same input features, then they will all provide the same output, and the random forest will be equivalent to a single decision tree. 

A reason that made Random Forests very popular is that they can be trained on data sets with more features than observations without prior feature selection, a characteristic that made then a relevant tool, for example, for gene expression problems in Bioinformatics. For more details, see~\cite{okun2007}.  

\section{Nuclear mass models}\label{nuc:m:model}

Before applying the decision tree to a more realistic case, we now introduce the nuclear models we are going to study.
According to the most recent nuclear mass table~\cite{wang2017ame2016}, more than 2000 nuclei have been observed experimentally. To interpret such a quantity of data, several mass models have been developed over the years~\cite{sob14} with various levels of accuracy. For this guide, we selected two simple ones: the Bethe-Weizs\"acker mass formula~\cite{krane1987introductory} based on the liquid drop (LD) approximation and the Duflo-Zuker~\cite{duf95} mass model.
The reason of this choice is twofold: the models contain quite different physical knowledge about the data, for example, the lack of shell effects in LD case, but they are relatively simple and not CPU intensive, thus giving us the opportunity to focus more on the statistical aspects of the current guide.

\subsection{Liquid drop}

\noindent Within the LD model, the binding energy ($B$) of a nucleus is calculated as a sum of five terms as a function of the total number of neutrons ($N$) and protons ($Z$) as

\begin{eqnarray}\label{bene}
B^{LD}_{th}(N,Z)=a_v A-a_sA^{2/3}-a_c\frac{Z(Z-1)}{A^{1/3}}-a_a\frac{(N-Z)^2}{A}-\delta\frac{\text{mod}(Z,2)+\text{mod}(N,2)-1}{A^{1/2}}\;,
\end{eqnarray}

\noindent where $A=N+Z$. The set of optimal parameters have been tuned in Ref.~\cite{pas19}. These parameters are named volume ($a_v$), surface ($a_s$), Coulomb ($a_c$), asymmetry ($a_a$) and pairing ($\delta$) and they refer to specific physical properties of the underlying nuclear system~\cite{krane1987introductory}.

\subsection{Duflo-Zuker}

The Duflo-Zuker \cite{duf95}  is a  macroscopic mass model based on a generalised LD plus the shell-model monopole Hamiltonian and it is used to obtain the binding energies of nuclei along the whole nuclear chart with quite a remarkable accuracy.
The nuclear binding energy  for a given nucleus is written as a sum of ten terms as

\begin{eqnarray}\label{beneDZ}
B^{DZ10}_{th}=a_1V_C+a_2 (M+S)-a_3\frac{M}{\rho}-a_4V_T+a_5V_{TS}+a_6s_3-a_7\frac{s_3}{\rho}+a_8s_4+a_{9}d_4+a_{10}V_P\;.
\end{eqnarray}

\noindent We defined $2T=|N-Z|$  and $\rho=A^{1/3}\left[ 1-\frac{1}{4}\left(\frac{T}{A}\right)^2\right]^2$.
The ten different contributions can be grouped into two categories: in the first one we find terms similar to the LD model as Coulomb ($V_C$), symmetry energy ($V_T,V_{TS}$) and pairing $V_P$. The other parameters originate from the averaging of shell-model Hamiltonian. See~\cite{men14} for more details. The model described in Eq.\ref{beneDZ} is usually referred as DZ10 and its parameters have been recently tuned in~\cite{pas19b}. Within the literature, it is also possible to find other versions with extra parameters~\cite{qi2015theoretical}, but we will not consider them here for the sake of simplicity.

\begin{figure*}
\begin{center}
\includegraphics[width=0.4\textwidth,angle=0, trim={0 30 0 0}]{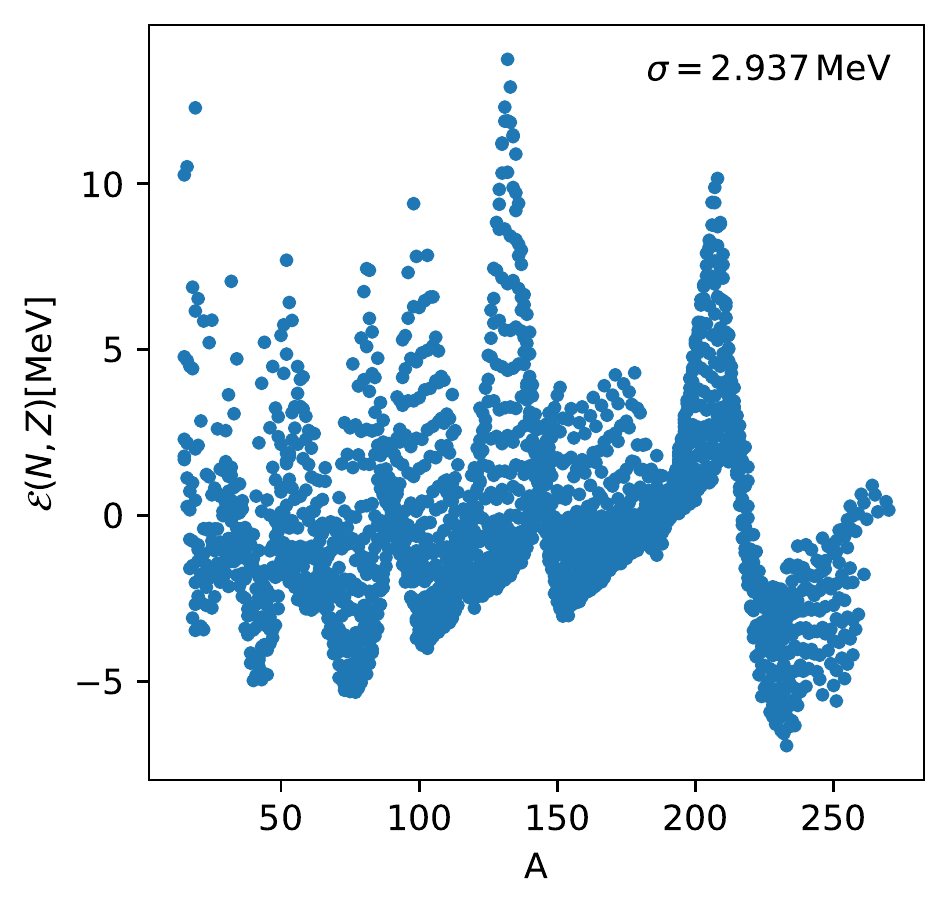}
\includegraphics[width=0.4\textwidth,angle=0, trim={0 30 0 0}]{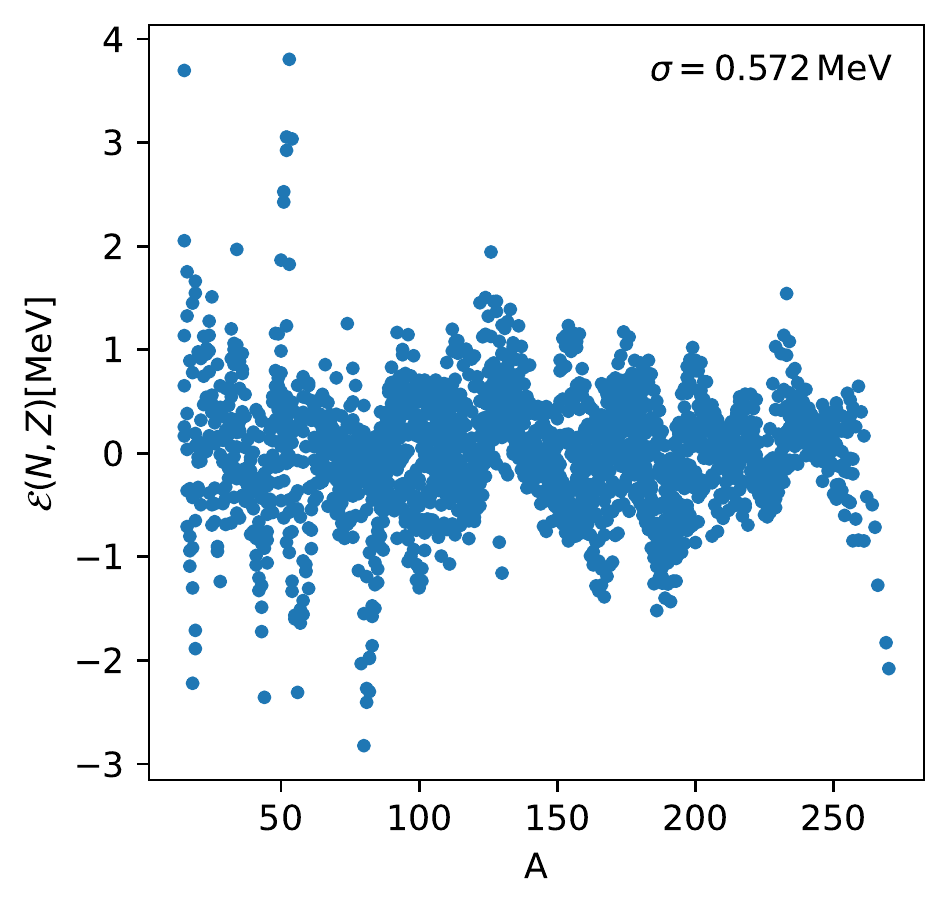}
\end{center}
\caption{Residuals (expressed in MeV) obtained using the liquid drop model (left panel) and the DZ10 model (right panel).}
\label{fig:res}
\end{figure*}

In Fig.\ref{fig:res}, we illustrate the behaviour of the residuals  $\mathcal{E}(N,Z)$ obtained with the two mass models \emph{i.e.} the difference between the empirical data and the models predictions. We assume that nuclear data~\cite{wang2017ame2016} have negligible experimental error compared to the model, and we discard all data having an uncertainty larger than 100 keV. This is a reasonable assumption to be made since the typical discrepancy between models and data is usually one or two orders of magnitude larger than the experimental errors~\cite{sob14}. See discussion in~\cite{pas19} for more details.

In each panel of Fig.\ref{fig:res}, we also provide the root mean square (RMS) deviation $\sigma$. We thus see that DZ10 is roughly one order of magnitude more accurate in reproducing data than the simple LD.
The detailed analysis of these residuals has been already performed in~\cite{pas19,pas19b} showing that they do not have the form of a simple white noise, but they contain a degree of correlation. 


\section{Results}\label{sec:res}

 We apply the decision tree to the case of nuclear data. Using the same notation adopted in the previous examples, the input $X$ is now a matrix with 3 columns $N,Z,A$ while the response $\hat{Y}$ is the residual.

As specified before, the goal is to minimise the RMS on unseen data or, in other words, learning without overfitting. While it appears obvious that a tree with only one leaf, which means replacing all the values of $Y$ with the average $\overline{Y}$, or with as many leaves as there are observations are not very useful, determining the optimal value for the number of leaves is not straightforward. The approach is empirical: experimenting with a reasonable set of values for the number of leaves, and pick the best results according to the cross-validation.  

With only one adjustable hyper-parameter like the maximum number of leaves, exploring the parameter space is straightforward: all the possible values are tested, with a cost of $M$ cross-validated models, where $M$ is the number of possible values for the number of leaves. In our example, exploring trees with a maximum number of leaves between 2 and 500 implies cross-validating for $M=499$ models. 

On the other hand, with regressors with many adjustable parameters, as for example Xgboost~\cite{chen2016}, exploring the hyper-parameter space is more challenging. For example, with 10 hyper-parameters, exploring $M$ values for each of them means exploring a grid with $10^{M}$ points. In this case, it is better to use dedicated libraries~\cite{komer2014}.

\begin{figure}
\begin{center}
\includegraphics[width=0.8\textwidth,angle=0, trim={0 30 0 0}]{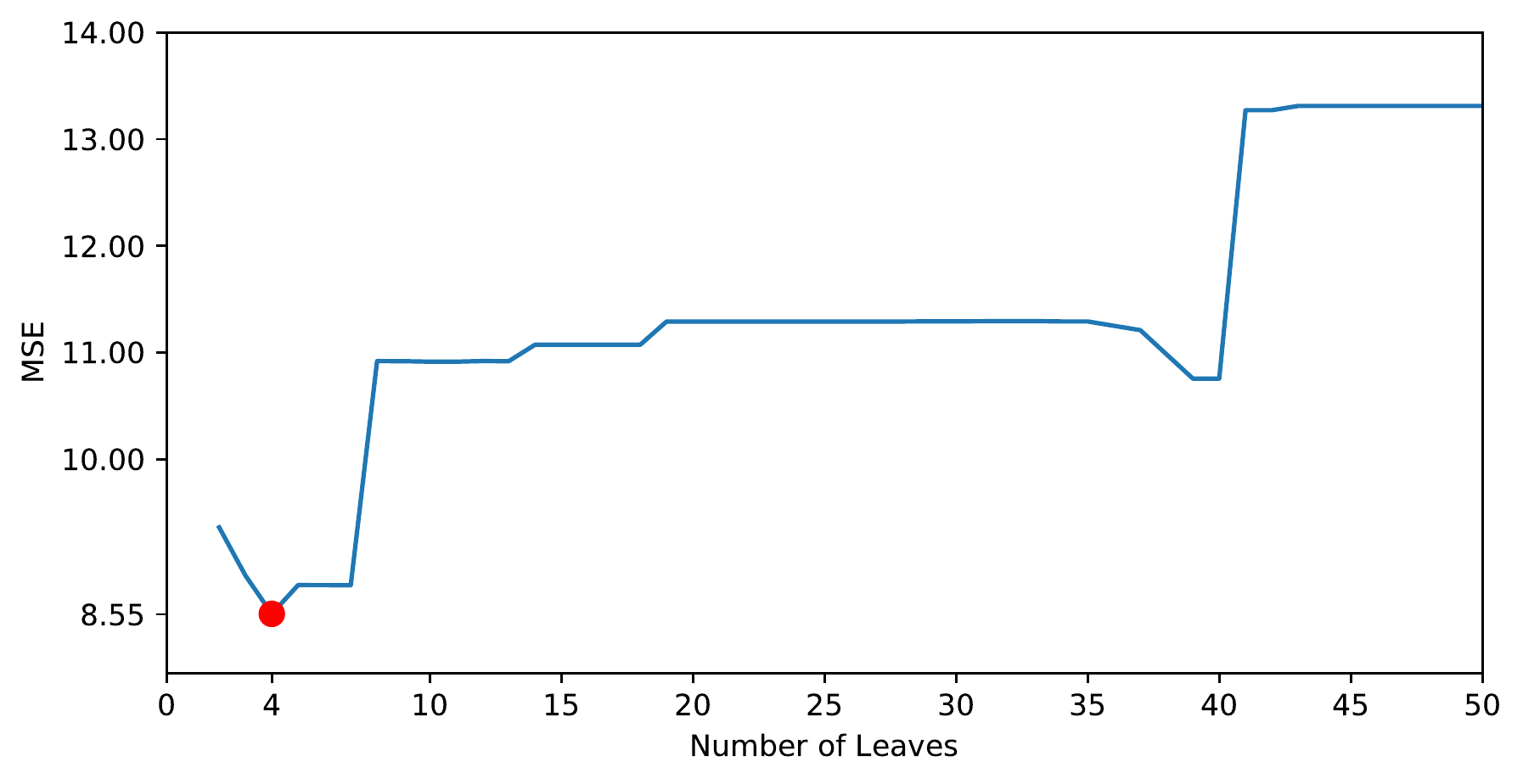}
\end{center}
\caption{Evolution of the MSE as a function of the number of leaves. The dot corresponds to the absolute minimum. See text for details.}
\label{fig:tree:MSE}
\end{figure}

As a first application, we train a simple decision tree over the residuals of the LD model as shown in the left panel of Fig.\ref{fig:res}. In Fig.\ref{fig:tree:MSE}, we illustrate the evolution of the MSE as a function of the number of leaves. For sake of clarity we truncated the figure to 50 leaves, the full plot can be found in the Supplementary Material.

\begin{figure*}[!h]
\begin{center}
\includegraphics[width=0.35\textwidth,angle=0, trim={0 30 0 0}]{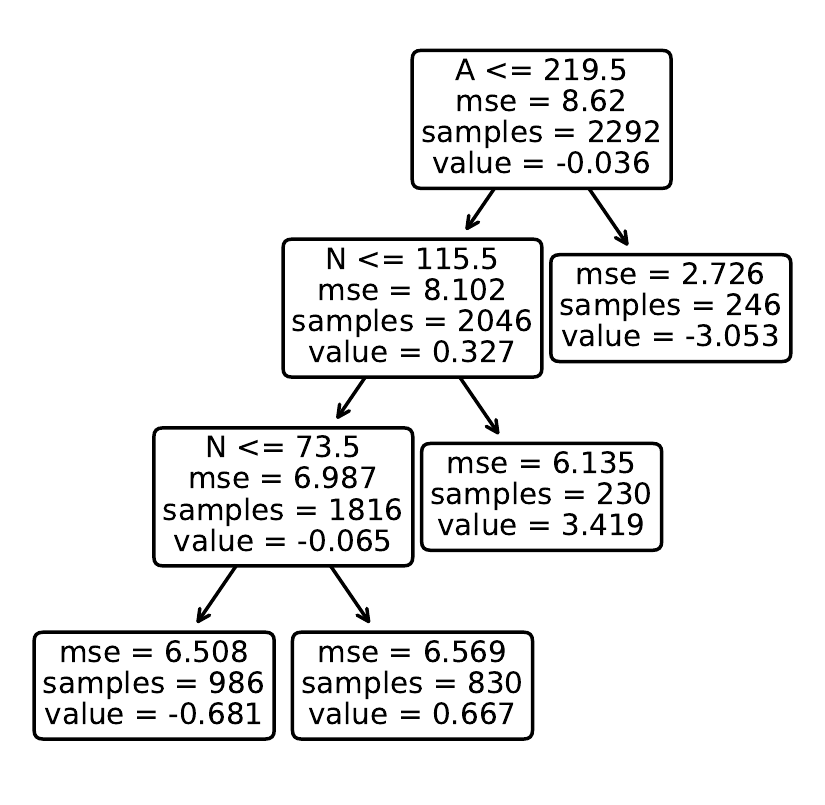}
\end{center}
\caption{Decision tree for LD model using only three features  $N,Z,A$. See text for details.}
\label{fig:tree:LD1}
\end{figure*}

From Fig.\ref{fig:tree:MSE}, we notice that the optimal number of leaves is four. The structure of the tree is reported in Fig.\ref{fig:tree:LD1}. By inspecting the splits of the data, we notice that the main feature of the data is associated with the neutron number $N$. The tree splits the nuclei in super-heavy ($A>219$) and non-super-heavy. Then it further splits into very neutron-rich and not. Finally, the tree separates out the remaining nuclei into two groups: light and heavy.

\noindent Having the optimal tree, we now translate it into a simple code. Here we use Fortran, but any other computer language can be used with no difficulty.

\begin{verbatim}
! Example of decision tree for the LD model with 2 features

!...
!input: N,Z (neutron/protons) / integers
!output: BEtree (correction to binding energy) /real
!...
  A=N+Z
  if(A < 219.5)then
     if(N < 115.5)then
        if(N < 73.5) then
           BEtree=-0.681
        else
           BEtree=0.667
        endif
     else
        BEtree=3.419
     endif
  else
     BEtree=-3.053
  endif
...
\end{verbatim}

\noindent Using the previous code, we now calculate the nuclear binding energies as 

\begin{eqnarray}\label{eq:mod1}
\mathcal{B}_{th}=B_{th}^{LD}+B_{tree}\;,
\end{eqnarray}

\noindent where $B_{tree}$ represents the binding energy calculated with the decision tree. By comparing the predicted masses obtained with Eq.\ref{eq:mod1} with the experimental ones, we obtain an RMS of $\sigma_{C,tr}=2.467$ MeV. This is what is called \emph{training error}, which is the RMS between the response and the predictions of the model trained on all data. A more conservative estimation that should be preferred is the validation error on unseen data, \emph{i.e} the RMS estimated on data that were not used during the training. In this case, $\sigma_{C,val} = 2.925$ MeV.

\begin{figure*}
\begin{center}
\includegraphics[width=0.8\textwidth,angle=0, trim={0 25 0 0}]{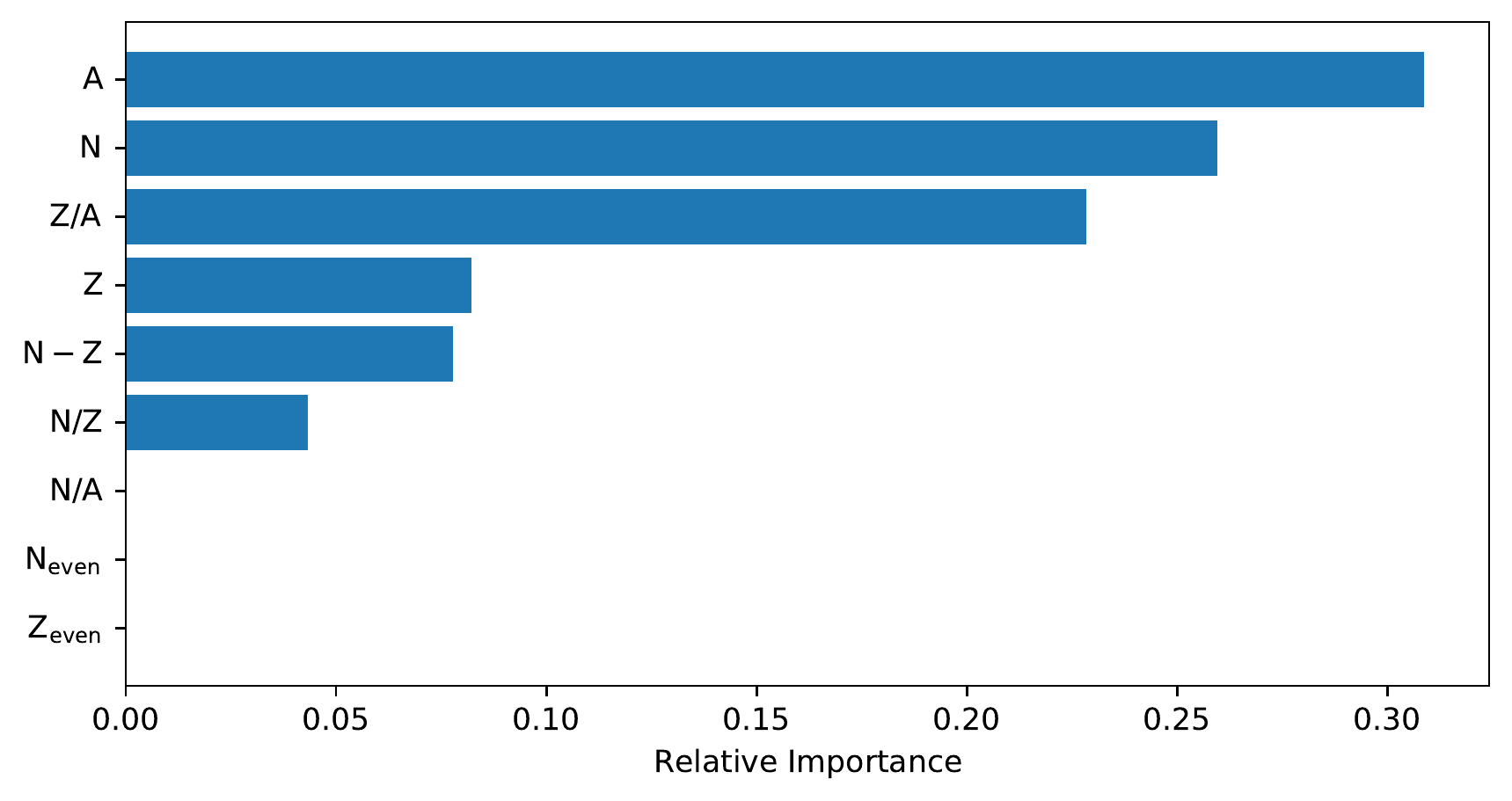}
\end{center}
\caption{Relative importance of the features (reduction in impurity normalised) in the Liquid Drop model. The features $N/A$, $N_{\text{even}}$ and $Z_{\text{even}}$ (equal to 1 if $N$ or $Z$ is even and 0 otherwise)  were not used in the model and as a consequence they have zero importance.}
\label{fig:tree:LD3}
\end{figure*}

It is possible to further improve on this result, by using Feature Engineering as discussed previously. To this respect, we provide some additional information to the tree: $A$, $N-Z$, $N/Z$, $Z/A$, $Z/N$ \dots. The full list of features can be seen from Fig.\ref{fig:tree:LD3}. By inspecting Eq.\ref{bene}, we observe that these features are already used to build the LD model and as such we help the decision tree to identify new patterns in the data.
It is worth noting here that other features may be used instead, but a monotonic transformation of existing features (like $A^{1/3}$ if we are using $A$) will provide little to no performance improvement. See for example \cite{heaton} for an empirical discussion of the topic. Identifying patterns into the data is of great help since it may lead (in complex cases) to better solutions.

\begin{figure*}
\begin{center}
\includegraphics[width=0.65\textwidth,angle=0, trim={0 25 0 0}]{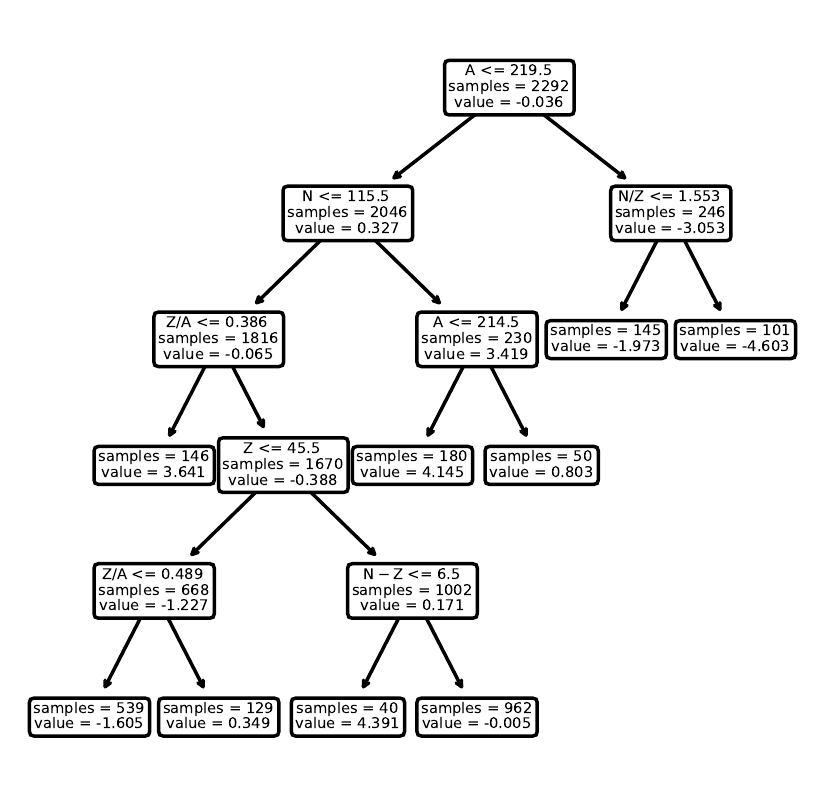}
\end{center}
\caption{Improved decision tree for LD model using Feature Engineering. For the sake of clarity and readability, the impurity (MSE) was omitted.}
\label{fig:tree:LD2}
\end{figure*}

In Fig.\ref{fig:tree:LD2}, we report the structure of this new tree. The optimal number of leaves is 9. By implementing this tree into a simple numerical code, as done previously, and applying it to the LD residuals we obtain a slight improvement. The total RMS over the entire nuclear chart now falls to $\sigma_{C,tr}=2.069$ MeV (on unseen data, $\sigma_{C,val} = 2.881$ MeV).

Although the decision tree performs less well (in terms of RMS) than a more complex neural network~\cite{utama2016nuclear}, we can still use it to identify possible trends in the data set. By inspecting Fig.\ref{fig:tree:LD3}, we observe that not all the 9 features have been used to build the code. In Fig.\ref{fig:tree:LD3}, we illustrate the relative importance of the features of the LD model, calculated using Eq.\ref{eq:importance}. We see that the proton fraction $Z/A$ is more important than the individual number of neutrons and protons. It is interesting to note that the decision tree is not affected by even/odd nuclei. This may imply that either the simple pairing term in Eq.\ref{bene} is enough to grasp the odd-even staggering, or the granularity required to the tree is too high, leading to a number of leaves comparable with the number of data points, or other features can surrogate the odd-even features.
We also observe that in this tree the total number of nucleons $A$ and the proton fraction $Z/A$ are as important or more than the number of neutrons $N$. This clearly explains why the performances of this new tree have improved compared to the one given in Fig.\ref{fig:tree:LD1}.
A detailed understanding of the trend in the data would require a more in-depth analysis, and so we leave it for future investigations.

\begin{figure*}
\begin{center}
\includegraphics[width=0.8\textwidth,angle=0, trim={0 30 0 0}]{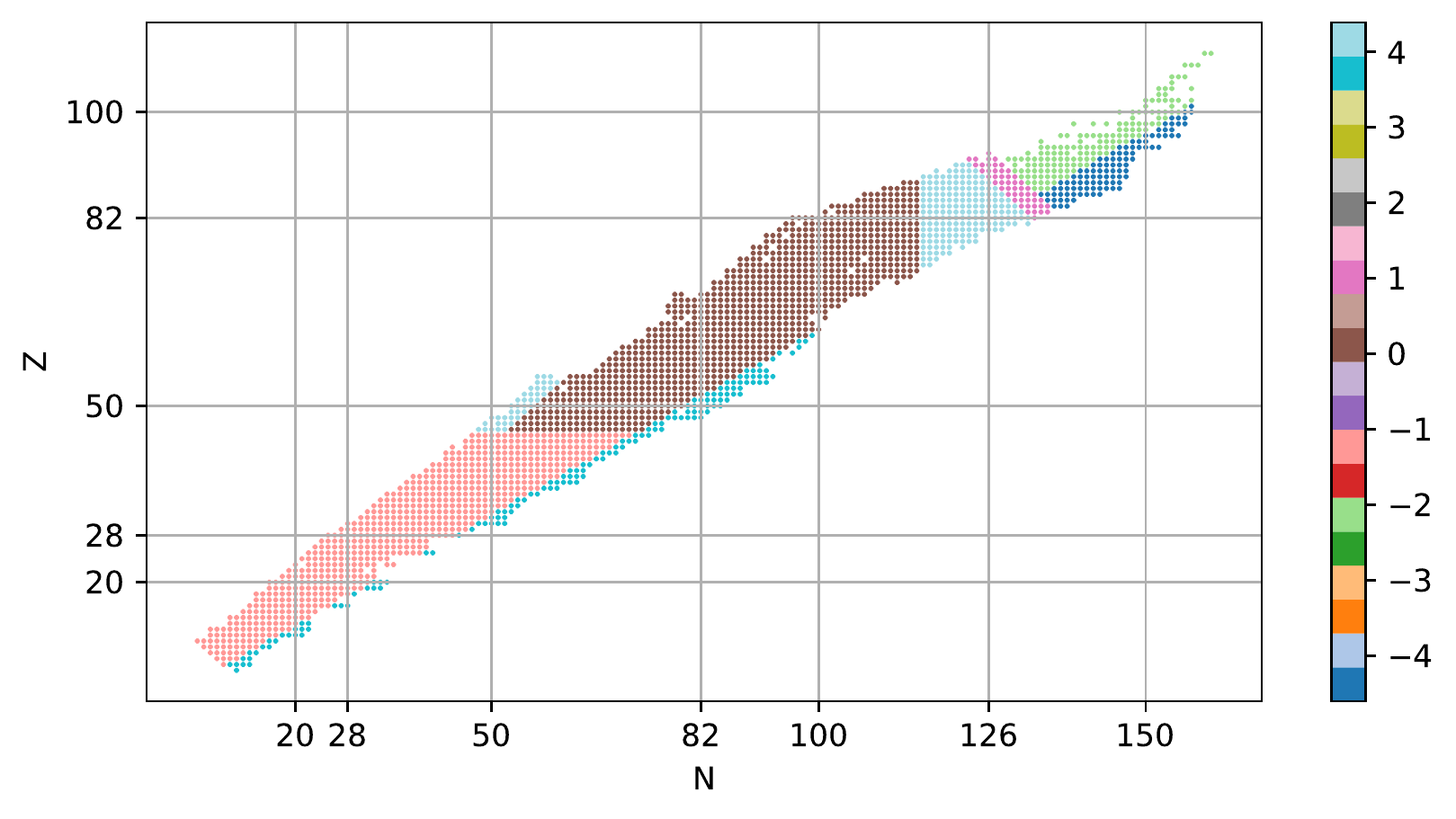}
\end{center}
\caption{(Colours online) Graphical representation of the splits done by the decision tree illustrated in Fig.\ref{fig:tree:LD2} on the Segr\'e chart of nuclei. The various zones correspond to the energy corrections expressed in MeV derived from the decision tree to the LD model as a function of $N$, $Z$.}
\label{fig:segre_ld}
\end{figure*}

In Fig.\ref{fig:segre_ld}, we present a graphical illustration of the energy corrections found by the decision tree for the different nuclei along the Segr\'e chart. This figure is an alternative way to represent the various leaves shown in Fig.\ref{fig:tree:LD2}.
We observe that we have 6 major splits along the valley of stability where we find light, medium-heavy and heavy nuclei. The latter are then still separated into 4 smaller groups.
The other cuts occur along the region of proton-rich and neutron-rich, thus the edges of the chart.
From this general overview, we may conclude that the residuals of the LD model are quite homogeneous (only two separations) along the valley of stability up to medium-heavy nuclei. Outside this range, the number of splits increase since the tree identifies a larger variation in the data. This may imply some missing physics in the model (choice of features) for these particular regions of the chart.

Having seen how the decision tree works for a schematic model as LD, we now apply it to the more sophisticated DZ10. We adopt the same features as shown in Fig.\ref{fig:tree:LD3} to obtain the best performances. Since the structure of the residuals is different there is no \emph{a priori} reason to use such features, but for the sake of simplicity of the current guide, we keep them the same. 

\begin{figure*}
\begin{center}
\includegraphics[width=0.7\textwidth,angle=0, trim={0 30 0 0}]{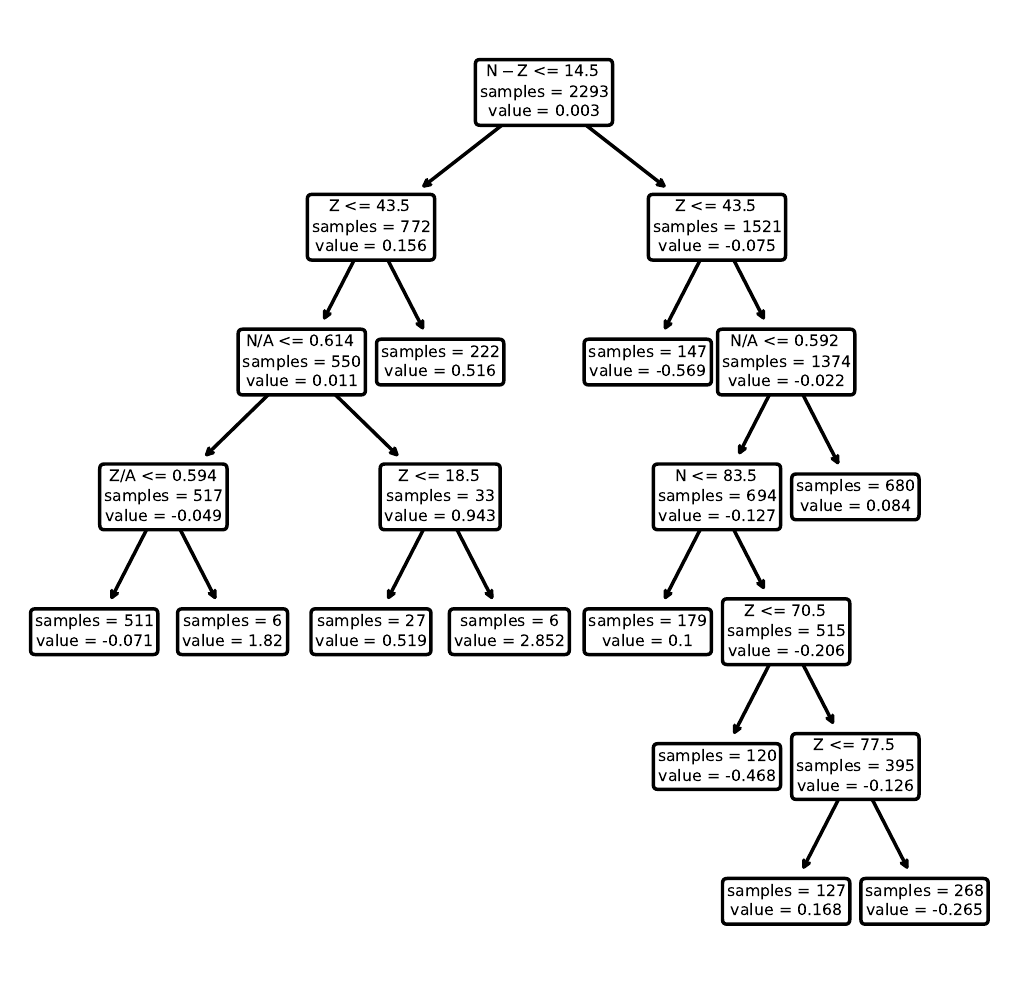}
\end{center}
\caption{Decision tree for  DZ10 model given in Eq.\ref{beneDZ}. For the sake of clarity and readability, the impurity (MSE) was omitted.}
\label{fig:tree:DZ}
\end{figure*}

For the DZ10  model, the optimal tree has now 11 leaves and it is illustrated in Fig.\ref{fig:tree:DZ}. As discussed previously, the tree can be easily translated into a small numerical code using a simple structure.

\begin{figure*}
\begin{center}
\includegraphics[width=0.8\textwidth,angle=0, trim={0 30 0 0}]{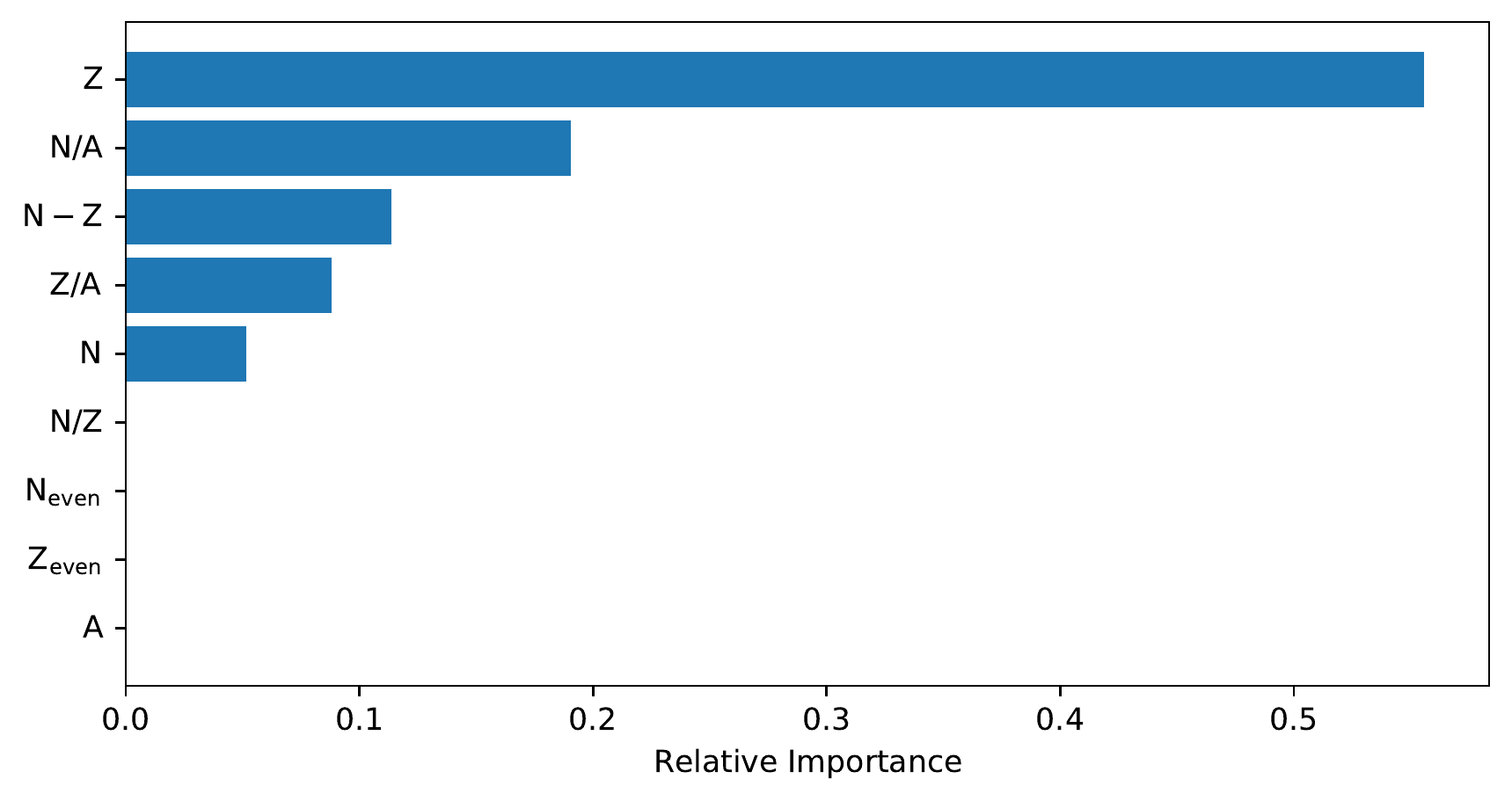}
\end{center}
\caption{Relative importance of the features (reduction in impurity normalised) in the DZ10 model. As before, the features $N/A$, $N_{\text{even}}$ and $Z_{\text{even}}$ (defined as in the caption of Fig.\ref{fig:tree:LD3}) were not used in the model, has zero importance in the model and can thus be discarded.}
\label{fig:tree:DZ10}
\end{figure*}

In Fig.\ref{fig:tree:DZ10}, we illustrate the importance of the features used to build such a tree. It is interesting to observe that the most important feature is the charge dependence and the isovector dependence of the model ($N-Z$). By comparing with Fig.\ref{fig:tree:LD3}, we observe that the relative importance of the features strongly depends on the model. In particular, four features of nine turned out not to be relevant during the optimisation of the tree.
We could further simplify the tree by reducing the features used or exploring new ones. This investigation goes beyond the scope of the present guide since we are only interested in illustrating how the algorithm works.

We implement such a tree within a simple Fortran code. See Appendix\ref{app:pseudo-code} for details.
With such a code, we calculate the new binding energies as

\begin{eqnarray}\label{eq:mod2}
\mathcal{B}_{th}=B_{th}^{\text{DZ10}}+B_{tree}\;.
\end{eqnarray}

\noindent The global RMS drops to $\sigma_{C,tr}=0.471$ MeV ($\sigma_{C,val}=0.569$ MeV). The improvement on the binding energies is not as good as the one obtained~\cite{pas19b} using a more complex neural network, but the model we produced is far simpler. It is worth noticing that the final model given in Eq.\ref{eq:mod2} is fully specified by 43 parameters (the 10 original parameters from the DZ10 models, the 7 pairs describing the variable and the value to split on, the 9 values of the response on each leaf).
This number is comparable with most nuclear mass models~\cite{liran1976semiempirical,moller1988nuclear,goriely2009skyrme,wang2011nuclear}, which perform similarly to Eq.\ref{eq:mod2}.

\begin{figure*}
\begin{center}
\includegraphics[width=0.8\textwidth,angle=0, trim={0 30 0 0}]{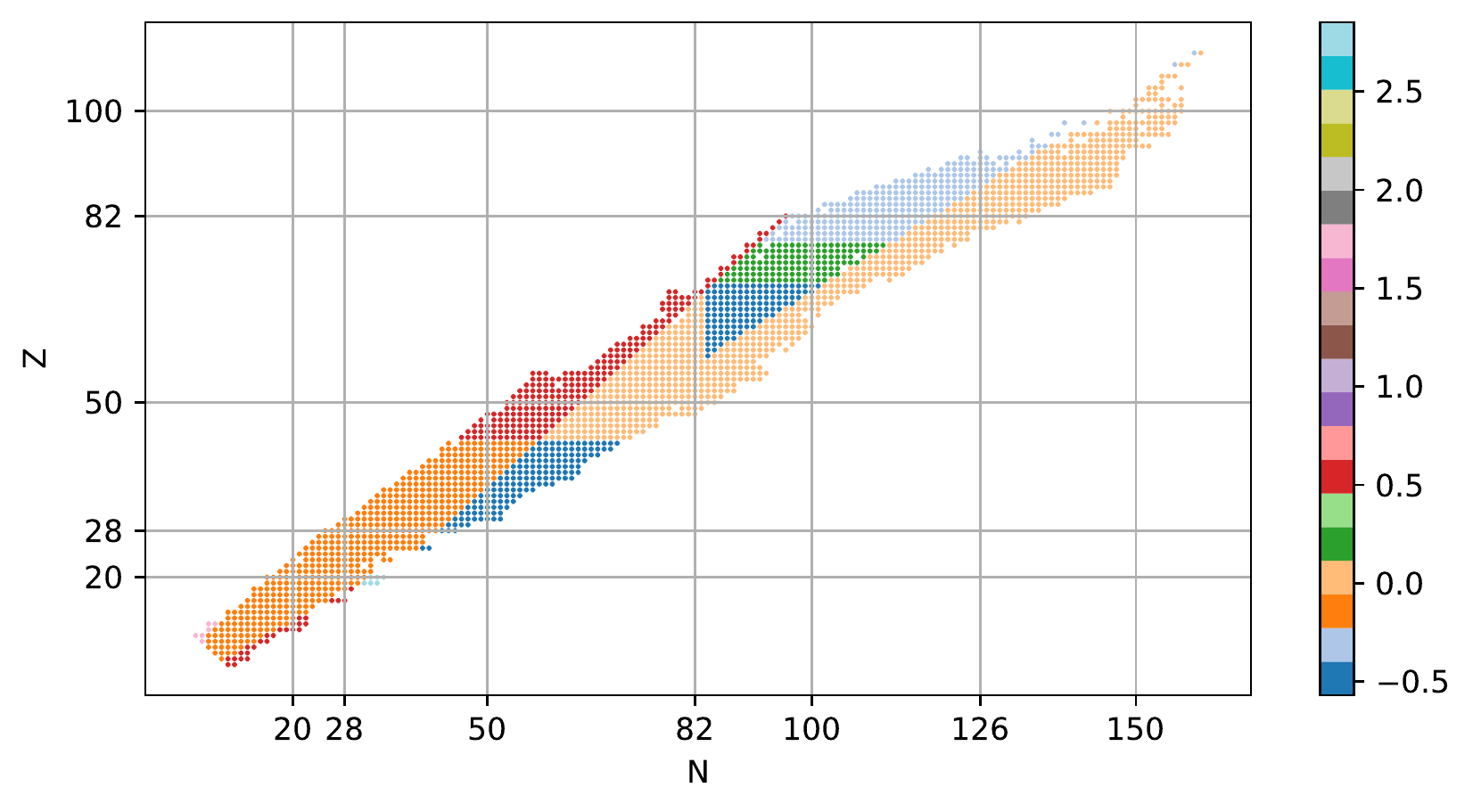}
\end{center}
\caption{(Colours online) Graphical representation of the splits done by the decision tree illustrated in Fig.\ref{fig:tree:DZ10} on the Segr\'e chart of nuclei. Corrections to the DZ10 model as a function of $N$, $Z$.}
\label{fig:segre_dz10}
\end{figure*}

As done previously for the LD model, we represent in Fig.\ref{fig:segre_dz10} the splits of the tree given in Fig.\ref{fig:tree:DZ}. We see that the most important cuts take place along $Z$. This was also the most important feature of the model  as shown in Fig.\ref{fig:tree:DZ10}. Interestingly, using the decision tree we have identified a large area in the residuals corresponding to the medium-heavy neutron-rich nuclei for which the correction is very small. On the contrary, the same mass range, but on the proton-rich side, requires a much more significant energy correction. This may be a symptom of poor treatment of the isovector channel in the model.

\begin{figure*}
\begin{center}
\includegraphics[width=0.8\textwidth,angle=0, trim={0 30 0 0}]{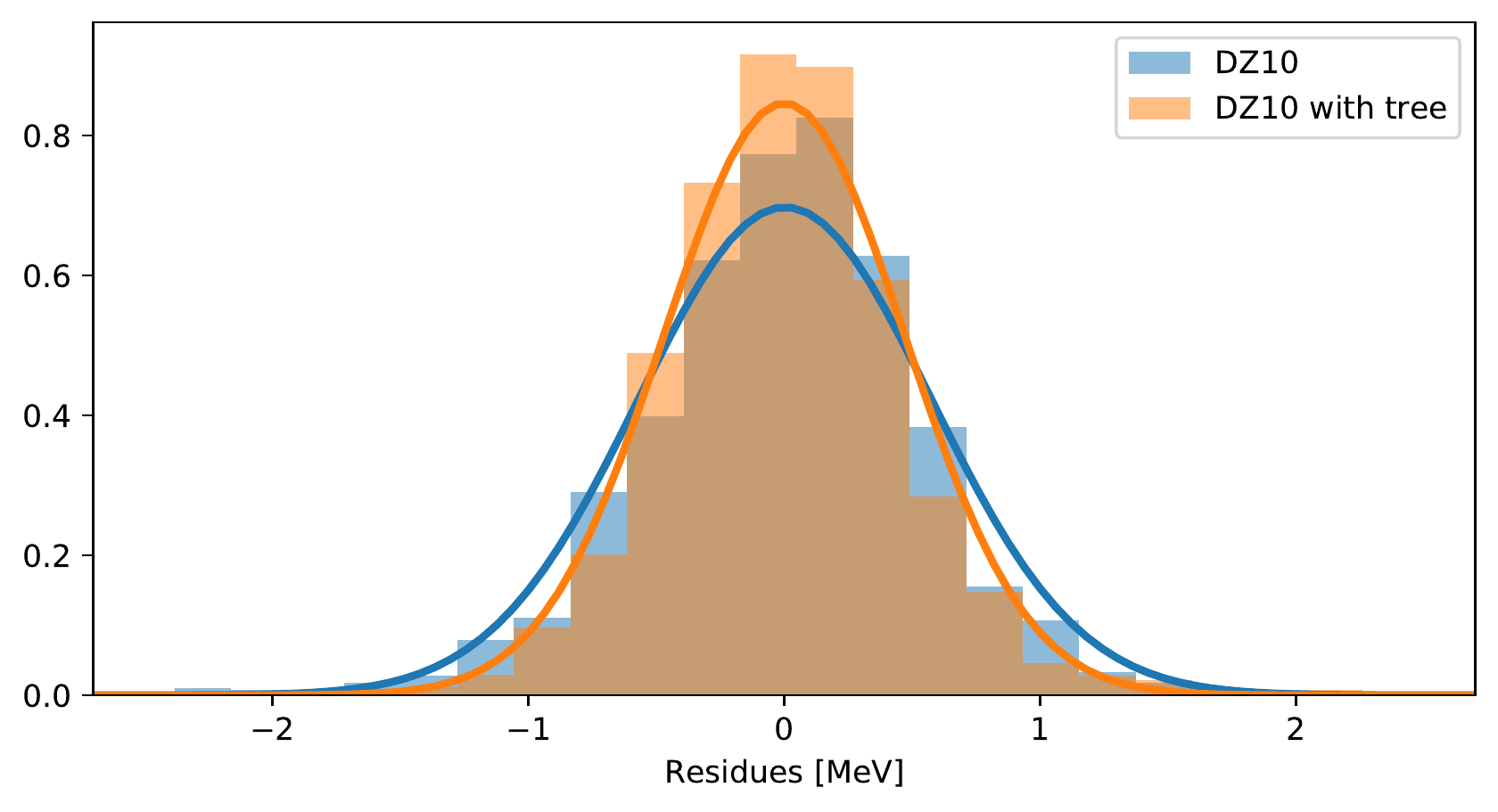}
\end{center}
\caption{(Colours online) Normalised histograms comparison between residues with and without the tree correction.}
\label{fig:histo}
\end{figure*}

In Fig.\ref{fig:histo}, we represent the comparison between the original residuals obtained with DZ10 model and the improved one using the decision tree. The histogram has been normalised. We see that the new residuals are now more clustered around the mean value, although we see that there are still some heavy tails that we have not been able to eliminate.
We have checked the normality of the residual using the standard Kolmogorov Smirnov test~\cite{barlow1993statistics} and we can say that the residuals are not normally distributed with a 95\% confidence, thus showing there is still some signal left in the data that we have not been able to grasp.

We conclude this section by summarising the impact of decision trees on the residuals of the various mass models and different features used in the calculations. The results are reported in Tab.\ref{tab:res}. We observe that using feature engineering, we have been able to reduce the RMS of the LD model by $\approx$30\%. Adopting the same features for the DZ10 model, we have improved the global RMS by $\approx18$\%.

\begin{table}
\begin{center}
\begin{tabular}{lcccc}
\toprule
Model & $\sigma_M$ & $\sigma_{C,tr}$ &  $\sigma_{C,val}$ & Improvement\\ \colrule
Liquid Drop \phantom{xxx} & 2.936 & 2.467 & 2.925 & 16.0\%\\
Liquid Drop With Features \phantom{xxx} & 2.936 & 2.070 & 2.881 & 29.5\% \\
DZ10 With Features \phantom{xxx} & 0.572& 0.471 & 0.569 &  17.6\% \\
\botrule
\end{tabular}
\end{center}
\caption{Here, $\sigma_M$ is the original model RMS, $\sigma_{C,tr}$ is the RMS once the corrections are added  $\sigma_{C,val}$ the RMS on unseen data (with corrections). \label{table:results}}
\label{tab:res}
\end{table}

It is worth noting that the numbers given in Tab.\ref{tab:res} are strictly dependent on the features we used to build the trees. Different choices would lead to different numbers.

\section{Conclusion}\label{sec:conc}

In this guide, we have illustrated a well-known decision tree algorithm by providing very simple and intuitive examples.
We have also shown the importance of analysing the data to improve the performances of the method.

We have applied the decision tree to the case of two well known nuclear mass models: liquid drop and Duflo-Zuker. In both cases, using a small number of leaves (9 and 11 respectively), we have been able to improve the global RMS of the models by 29.5\% and 17.6\% respectively. More consistent improvements have been obtained in the literature using neural networks~\cite{pas19b,utama2017refining,neufcourt2018bayesian}, but using a larger set of adjustable parameters.

We have also illustrated how to represent graphically the decision tree to better highlight the regions of the splits: this allows us to identify possible patterns in the data-set and eventually use them to improve the original model. By analysing the importance of the features, it is then possible to identify possible missing structure in the model

Finally, we have also illustrated how to translate the decision tree into a simple numerical code that could be easily added to existing ones to calculate nuclear masses.

\section*{Acknowledgments}

We thank M. Shelley for helping us with this guide. This work has been supported by STFC Grant No. ST/P003885/1.


\begin{appendix}
\section{Decision tree: pseudo-code}\label{app:pseudo-code}

For completeness, we provide here a possible translation of the decision tree into a Fortran code.

 \begin{verbatim}
!
! Example of decision tree for DZ10 model 
!

!input: N,Z (neutron/protons) / integers
!output: BEtree (correction to binding energy) /real

...

  if(N-Z< 14.5)then
     if(Z < 43.5)then
        if(N/A <= 0.614)then
           if(Z/A <= 0.594)then
              BEtree=-0.071
           else
              BEtree=1.82
           endif
        else
           if(Z < 18.5)then
              BEtree=0.519
           else
              BEtree=2.852
           endif
        endif
     else
        BEtree=0.516
     endif
  else
     if(Z < 43.5)then
        BEtree=-0.569
     else
        if(N/A <= 0.592)then
           if(N < 83.5)then
              BEtree=0.1
           else
              if(Z < 70.5)then
                 BEtree=-0.468
              else
                 if(Z < 77.5)then
                    BEtree=0.168
                 else
                    BEtree=-0.265
                 endif
              endif
           endif
        else
           BEtree=0.084
        endif
     endif
  endif
 \end{verbatim}
 
 Notice that a decision tree is formed by a simple sequence of conditional statements and the example given here can be easily ported to any other used computer language.
\end{appendix}

\bibliography{biblio}

\end{document}